\documentclass[aps,prl,twocolumn,groupedaddress]{revtex4}
\usepackage{graphicx}
\begin{document}

\title{Understanding the problem of glass transition on the basis of elastic waves in a liquid}
\author{Kostya Trachenko$^{1}$}
\author{V. V. Brazhkin$^{2}$}
\address{$^1$ Department of Earth Sciences, University of Cambridge, Cambridge CB2 3EQ, UK}
\address{$^2$ Institute for High Pressure Physics, RAS, 142190, Troitsk, Moscow Region, Russia}
\begin{abstract}
We propose that the properties of glass transition can be understood on the basis of elastic waves. Elastic waves originating from atomic jumps in a liquid propagate local expansion due to the anharmonicity of interatomic potential. This creates dynamic compressive stress, which increases the activation barrier for other events in a liquid. The non-trivial point is that the range of propagation of high-frequency elastic waves, $d_{\rm el}$, increases with liquid relaxation time $\tau$. A self-consistent calculation shows that this increase gives the Vogel-Fulcher-Tammann (VFT) law.
In the proposed theory, we discuss the origin of two dynamic crossovers in a liquid: 1) the crossover from exponential to non-exponential and from Arrhenius to VFT relaxation at high temperature and 2) the crossover from the VFT to a more Arrhenius-like relaxation at low temperature. The corresponding values of $\tau$ at the two crossovers are in quantitative parameter-free agreement with experiments. The origin of the second crossover allows us to reconcile the ongoing controversy surrounding the possible divergence of $\tau$. The crossover to Arrhenius relaxation universally takes place when $d_{\rm el}$ reaches system size, thus avoiding divergence and associated theoretical complications such as identifying the nature of the phase transition and the second phase itself.
Finally, we discuss the effect of volume on $\tau$ and the origin of liquid fragility.

\end{abstract}

\maketitle

\section{Introduction}

The problem of glass transition has been widely discussed \cite{angell,dyre,phil,langer,nat,kivel,angell1}, and has been considered as one of the deepest and most interesting challenges in physics \cite{ander}. As widely perceived, a glass transition theory should provide a consistent explanation of several universal properties of liquids which set in on lowering the temperature, including the physical origin of the Vogel-Fulcher-Tammann (VFT) law, slow non-exponential relaxation and dynamic crossovers \cite{angell,dyre,phil,langer,nat,kivel,casa,cross1,roland,cross2,sti,sti1,angell1}. The most widely studied property is the unusual behaviour of liquid relaxation time, $\tau$. On lowering the temperature, $\tau$ is almost never Arrhenius, but follows the VFT law: $\tau=\tau_0\exp\left(\frac{A}{T-T_0}\right)$, where $A$ and $T_0$ are constants \cite{angell,dyre,phil,langer,nat,kivel,angell1}.

As recently reviewed \cite{dyre}, the quest to understand the origin of the VFT law and other anomalous features of glass transition has resulted in the development of many theories and models, which discuss different parameters that control glass transition: free volume, entropy, energy landscape, mode coupling and others. However, there is no agreement as to what physical parameter governs glass transition \cite{dyre}. For this and other reasons, it has been proposed that glass transition remains a mystery, with no simple picture emerging \cite{angell,dyre,phil,langer,nat,kivel,angell1}.

Existing theories are often elaborate, and approach glass transition as an outstanding phenomenon that requires novel or special ideas and mechanisms \cite{dyre}. Yet we feel that one should be able to describe the process of cooling a liquid to a glass using familiar physical concepts which may, however, operate in a non-trivial and unexpected way. We suggest that a property relevant to glass transition is elasticity because a glass differs from a liquid only by its ability to support static shear stress. Hence, we approach the problem by asking whether glass transition can be understood 
on the basis of liquid elastic properties. 

Elastic approaches to glass transition were discussed previously (see, e.g., Ref. \cite{dyre,novikov,nemilov}), but the problem of explaining glass 
transition from the first principles remains. Consequently, there is no microscopic understanding of the origin of the VFT law, slow relaxation, 
dynamic crossovers and other effects of glass transition.

Recently, we proposed that several important properties of glass transition can be understood on the basis of elastic waves in a liquid \cite{our1,our2}. We considered the case when a liquid is perturbed and relaxes to equilibrium. However, it is important to consider elastic waves that originate from local atomic jumps in an equilibrium liquid. The main question is why and how these waves can result in the slowing down of liquid dynamics, the VFT law and dynamic crossovers.

In this paper, we develop and extend our approach to glass transition. Considering several anomalous, yet universal relaxation laws discussed above, we explore
non-trivial and unusual ways in which liquid elastic properties may emerge during glass transition. We analyze elastic waves originating from atomic
jumps in a liquid, and find that their effect is to create a dynamic compressive stress which slows down relaxation of other events. The non-trivial point is that the range of propagation of these waves increases with liquid relaxation time. A self-consistent calculation shows that this increase gives the VFT law. In the proposed theory, we discuss the origin of two dynamic crossovers in a liquid, the absence of divergence of $\tau$ at $T_0$, the effect of volume on $\tau$ and the origin of liquid fragility.

\section{Elastic interaction between local relaxation events}

Unlike in solids, atoms in liquids are not fixed in space, but are constantly rearranging. This gives liquid flow. Each flow event is a jump of an atom from its surrounding ``cage'', accompanied by large-scale rearrangement of the cage atoms. We call this process a local relaxation event (LRE). A LRE lasts on the order of Debye vibration period $\tau_0\approx$0.1 ps.

There are two known basic properties of LREs. The first property concerns liquid relaxation time, $\tau$. $\tau$ was phenomenologically introduced by Maxwell in the viscoelastic picture of flow as $\tau=\eta/G_{\infty}$, where $\eta$ is liquid viscosity and $G_{\infty}$ is the instantaneous shear modulus \cite{max}. Frenkel offered microscopic interpretation of $\tau$ as the time between LREs at one point in space in a liquid \cite{frenkel}. At high temperature, $\tau\approx\tau_0$. When $\tau$ increases to $\tau\approx 10^3$ s at glass transition temperature $T_g$, a liquid, by convention, forms a glass \cite{dyre}. The second property is that a LRE requires an increase of local volume. As widely discussed \cite{dyre,frenkel}, the activation barrier for a LRE at constant cage volume is very large due to strong short-range interatomic repulsions. Hence, atoms in the cage need to increase its volume in order to allow for the escape of the central atom (see Fig 1a). In doing so, work is performed to deform the surrounding liquid. This probes liquid elasticity. The work against the elastic force is equal to the activation barrier for a LRE, $U$ \cite{dyre,frenkel}. This barrier is surmounted by temperature fluctuations, so that $\tau=\tau_0\exp(U/T)$ ($k_{\rm B}=1$) \cite{frenkel}.

\begin{figure}
\begin{center}
{\scalebox{0.55}{\includegraphics{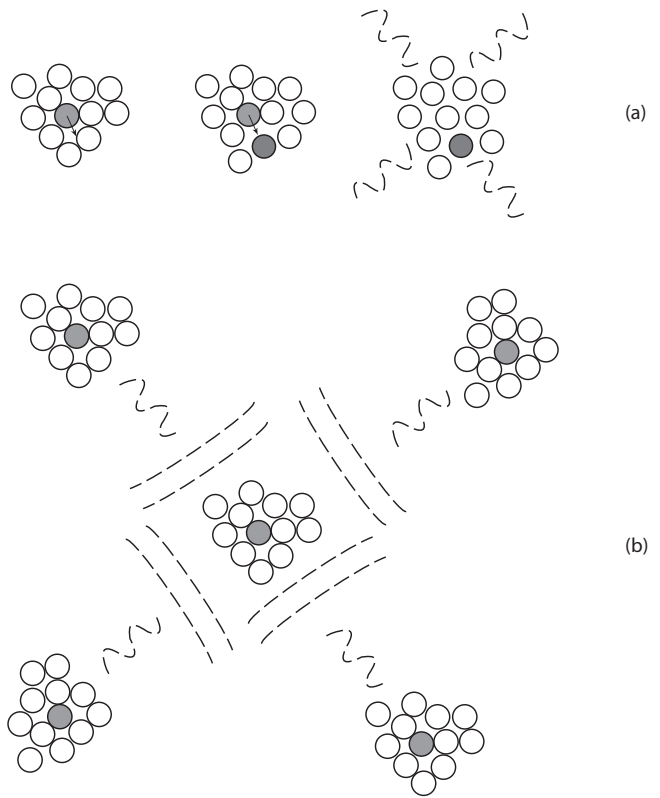}}}
\end{center}
\caption{(a) Large-scale cage restructuring due to the atomic jump induces a propagating high-frequency wave. This wave propagates volume expansion due to anharmonicity; (b) As a result of the arrival of compressive wave fronts, atoms in the central cage are under dynamic compressive stress.}
\end{figure}

An important insight into glass transition comes from the realization that LREs interact elastically, as we have recently proposed \cite{our1,our2}. A LRE involves restructuring of the cage that involves large-amplitude atomic motions of about of 1--2 \AA\ (see Fig. 1a). On the very short time scale of a LRE (when $\tau_0<\tau$), the surrounding liquid can be viewed as an elastic medium \cite{frenkel}. Therefore, the large atomic motion from a LRE elastically deforms the surrounding liquid, inducing elastic waves. Because their wavelengths are on the order of interatomic separations, the frequency of these waves, $\omega$, is on the order of Debye frequency. This means that in almost entire range of $\tau$ that is relevant for glass transition, $\omega>1/\tau$ holds true. As discussed by Frenkel, high-frequency $\omega>1/\tau$ waves are propagating in a liquid as in a solid \cite{frenkel}. The existence of propagating high-frequency waves in liquids is now firmly established: the conclusion from numerous experiments is that liquids support vibrational modes which extend down to wavelengths comparable to interatomic separations, similar to phonons in solids \cite{pilgrim}.

The propagating high-frequency waves from a LRE deform the cages around other atoms in a liquid. This affects their relaxation because, as discussed above, the jump of an atom depends on its cage. Therefore, LREs {\it interact} via the elastic waves they induce. Elastic interaction between LREs is the physical origin of {\it cooperativity} of relaxation in a liquid, whose physical origin has been much discussed, but remained unclear \cite{angell,dyre,langer,kivel}. The key issue is the range of this interaction.

Lets consider how a solid-like elastic wave is affected by LREs in a liquid. Because $\tau$ sets the period of structural rearrangements in a liquid, it defines the time of decay of induced static shear stress \cite{frenkel,landau} as well as of high-frequency ($\omega>1/\tau$) solid-like propagating waves \cite{dexter}. Note that $\tau$ is similar for high-frequency shear and longitudinal stress \cite{dexter}. If $c$ is the speed of sound, $d_{\rm el}=c\tau$ gives the length of stress decay:

\begin{equation}
d_{\rm el}=c\tau
\label{2}
\end{equation}

At the microscopic level, $d_{\rm el}=c\tau$ originates as follows. A high-frequency $\omega>1/\tau$ wave induced by a LRE propagates as in a solid until a remote LRE takes place at the wave front. At this point, atoms in the wave front do not pass the oscillations further on as in a solid with constant structure, but are involved in large scale structural fluctuation due to the LRE motion. Hence, the solid-like elastic wave is absorbed by the motion of the liquid-like atomic cluster due to the LRE. Suppose the remote LRE meets the wave front distance $d_{\rm el}$ away from the original LRE. $d_{\rm el}$ is defined from the equality of the wave travel time, $d_{\rm el}/c$, and the time at which the remote LRE takes place at point $d_{\rm el}$. The latter time is given by $\tau$ because microscopically, $\tau$ is the average time between LREs at one point in space, and we find $d_{\rm el}=c\tau$ as before.

Hence, $d_{\rm el}=c\tau$ is the distance over which a high-frequency wave propagates in a liquid as it would in a solid without its structure being modified by LREs. In an ideal crystal with infinite $\tau$, Eq. (\ref{2}) gives the infinite range of wave propagation, as expected.

It is interesting to note that $d_{\rm el}=c\tau$ is in agreement with Frenkel's theory of viscoelastic relaxation in a liquid. In this theory, waves in a liquid decay with distance $x$ as $\propto\exp(-x/d)$. Using the condition $\omega\tau>1$ explicitly, Frenkel's theory gives $d\approx c\tau$. This is discussed in detail in the Appendix.

We emphasize that similar to Frenkel's theory, our derivation of $d_{\rm el}=c\tau$ is also based on the condition $\omega\tau>1$, albeit implicitly. Indeed, we approached a liquid from the solid (elastic) phase, and considered how solid-like waves are affected by LREs in a liquid. This approach is therefore based on the assertion that solid-like elastic equilibrium exists in a liquid, which is the case for high-frequency $\omega>1/\tau$ waves.

The non-trivial point here is that $d_{\rm el}=c\tau$ {\it increases} with $\tau$. This is directly opposite to the usual decay of hydrodynamic waves, whose propagation range varies as $1/\tau$ (see also, Appendix). Crucially, this is because the considered solid-like elastic regime of wave propagation ($\omega\tau>1$) is markedly different from the commonly discussed hydrodynamic regime ($\omega\tau<1$). In the latter, LREs are frequent enough to eliminate the state of elastic equilibrium, and establish hydrodynamic equilibrium instead. The resulting equations of motion are those of hydrodynamics and viscous flow \cite{hydro}. Interestingly, these equations are widely used to describe liquid dynamics and glass transition, yet they are not applicable to our approach to glass transition based on high-frequency interactions. The presence of these interactions makes our approach essentially non-hydrodynamic, but elastic instead.

$d_{\rm el}$ can be called liquid elasticity length, because it defines the range over which two LREs interact with each other via induced high-frequency elastic waves. Importantly, $d_{\rm el}=c\tau$ {\it increases} on lowering the temperature because $\tau$ increases. We propose that this is the key to the problem of glass transition.

We finish this section with a comment regarding the generality of our discussion. Depending on liquid structure and interactions, LREs may take different form. For example, a LRE in covalent network liquids may involve bond switching from under- to over-coordinated states, whereas in spherically symmetric systems (e. g., ionic or metallic) it may resemble an illustration in Fig. 1. Hence, the way in which high-frequency waves are generated by 
LREs may be system-specific. However, the interaction of LREs via induced elastic waves is general, and should apply to all liquids
undergoing glass transition, including covalent, ionic, molecular, metallic, polymeric and others. Consequently, we expect that the VFT law, dynamic
crossovers and other effects of glass transition discussed below can be understood on the basis of elastic interactions and $d_{\rm el}$.

\section{Dynamic compressive stress}

Lets consider the nature of LRE-induced waves in more detail. In between LREs, atoms in a given local region vibrate with small amplitudes as in a solid (glass) \cite{frenkel}, and a harmonic approximation can be applied to the same extent as to the solid phase. As discussed above, a LRE involves large-scale atomic motions of 1--2 \AA. This considerably widens the distribution of interatomic separations $\Delta r$. At large $\Delta r$, harmonic approximation no longer applies, and the potential anharmonicity becomes important. In particular, the decrease of $\Delta r$ results in short-range repulsion that is always stronger than the attraction due to the same increase of $\Delta r$. Because expansion carries smaller energy penalty, large-amplitude motion of atoms involved in a LRE results in short-lived expansion of local volume around the LRE. In a simplified picture, this process can be thought of as the appearance of ``hot'' local regions in a liquid due to LRE motion and associated local thermal expansion.

Local volume expansion is propagated away from a LRE by high-frequency elastic waves discussed above. We note that when a sphere expands in a static elastic medium, no compression takes place at any point. Instead, the system expands by the amount equal to the increase of the sphere volume \cite{frenkel}, resulting in a pure shear deformation. The strain components $u$ from an expanding sphere (noting that $u\rightarrow$0 as $r\rightarrow\infty$) are $u_{rr}=-2b/r^3$, $u_{\theta\theta}=u_{\phi\phi}=b/r^3$ \cite{landau}, giving pure shear $u_{ii}=0$. As a result, the energy to statically expand the sphere of radius $r$ by amount $\Delta r$ depends on shear modulus $G$ only: $E=8\pi G r\Delta r^2$ \cite{frenkel}.

Unlike in the static case, there is compression at the front of the expanding wave. At time $t$, the wave front causes an outward displacement of atoms on the sphere of radius $ct$. This takes place during time $\tau_0$, i.e. very fast, because the wave is due to the remote LRE that lasts approximately $\tau_0$. Until this displacement causes the motion of the next concentric sphere with radius of $ct+x$, where $x$ is on the order of the interatomic separation, there exists a brief compressive stress in a layer of thickness $x$. This stress exists over time approximately equal to the period of vibrations of the system of two atoms, or $\tau_0$, i.e. is very short. As the wave propagates further, dynamic compression between the two layers disappears, and shear deformation is established as discussed above.

Lets now fix an atomic cage in the centre in Fig. 1b. As the front of the wave propagating volume expansion from a remote LRE arrives at the centre, it causes a brief compressive stress between the cage boundary atoms and the atoms in the next layer that include the central atom (see Fig. 1b). Hence, it puts the cage atoms under {\it dynamic compressive stress} (DCS). Because, as discussed above, the jump of the central atom requires cage expansion, the compressive wave fronts arriving at the centre result in more work needed to expand the cage. This increases the activation barrier for the central LRE, $U$, and slows down liquid dynamics.

We emphasize that DCS in a liquid is created by new elastic waves, which are notably absent in solids, and which propagate local volume expansions from the hot (in the sense discussed above) local regions. This process can be compared to pressure waves generated by local laser heating of a solid. In this case, the energy that heats up local regions and creates these pressure waves is external. In a liquid, the energy to create DCS is internal, i.e. it is the liquid's thermal energy. Here, as temperature increases and the liquid state is approached from the solid phase, LREs appear as a new type of ``hot'' local motion. This motion gives rise to a new set of expanding elastic waves, and DCS emerges as a result.

\section{Derivation of the VFT law}

The stage is now set for the calculation of $U$. $U$ is equal to the total work required to expand the cage by the amount required for a LRE to take place \cite{frenkel}. $U$ can be written as $U=U_0+U_1$. Here, $U_0$ is the high-temperature activation barrier that depends on liquid interatomic
forces and structure, but not on the elastic interactions with other LREs, i.e. is non-cooperative, or intrinsic. $U_1$ represents the interaction 
(cooperative) term. $U_1$ is equal to the additional work to expand the cage due to the arriving compressive wave fronts from other LREs, and is set 
dynamically. Let $q$ be the increase of the cage volume required for a LRE to take place. $q\approx a^3$, where $a$ is the interatomic separation of 
about 1 \AA. As discussed above, DCS exists in the arriving wave front, in a layer of thickness $a$. Lets consider DCS created by a remote LRE $i$ 
distance $r$ away from the centre in Figure 1b. If $p_i(r)$ is the value of DCS at the centre, the remote LRE contributes $qp_i(r)$ to $U_1$. Then, $U_1=\sum\limits_{i=1}^{N_{\tau}}qp_i(r)$, where $N_{\tau}$ is the number of compressive wave fronts that pass through the centre during time $\tau$, and $U$ reads:

\begin{equation}
U=U_0+\sum\limits_{i=1}^{N_{\tau}}qp_i(r)
\label{1}
\end{equation}

According to Eq. (\ref{2}), the sum in Eq. (\ref{1}) includes elastic waves from LREs inside the sphere of radius $d_{\rm el}=c\tau$. Importantly, each of the local relaxing regions inside this sphere contributes to the sum once. This is due to two reasons. First, because the central event relaxes during time $\tau$, waves from all the events located distance $c\tau$ away from the centre have enough time to propagate to the central point. Second, because a remote event also relaxes during time $\tau$, it contributes only one wave during the time of relaxation of the central event. Hence, the introduced length $d_{\rm el}=c\tau$ is self-consistent in that it accounts for the dynamical nature of LREs as well as for the wave dissipation.

Therefore, Eq. (\ref{1}) can be written as

\begin{equation}
U=U_0+q\int\limits_{d_0/2}^{d_{\rm el}} 4\pi\rho r^2 p_i (r) {\rm d}r
\label{3}
\end{equation}

\noindent where $\rho=\frac{6}{\pi d_0^3}$ is the density of local relaxing regions and $d_0$ is the region diameter of about 10 \AA.

$p_i (r)$ decreases with $r$. Recall that the short-lived DCS from a LRE originates at time $t$ in a thin layer of thickness $x$, when the outward motion of atoms takes place in a sphere of radius $ct$, but not yet in a sphere of radius $ct+x$. The strain and stress on the sphere of radius $ct$ decay as in elastic medium. This is because the waves under consideration are of high frequency ($\omega>1/\tau$) and therefore propagate in elastic equilibrium as discussed above. In elastic medium, the strain on an expanded sphere (i.e. on the sphere of radius $ct$) decays as $u\propto\frac{1}{r^3}$ \cite{landau}. This strain creates $p_i$ which, therefore, also decays as $p_i(r)\propto\frac{1}{r^3}$. Let $p_0$ be the value of $p_i$ at the cage boundary, distance $d_0/2$ away from the centre of a relaxing region. Then, $p_i(r)=p_0\left(\frac{d_0}{2r}\right)^3$, and Eq. (\ref{3}) becomes

\begin{equation}
U=U_0+3qp_0\ln\left(\frac{2d_{\rm el}}{d_0}\right)
\label{4}
\end{equation}

According to Eq. (\ref{4}), $U$ increases with $d_{\rm el}$. Because $d_{\rm el}=c\tau$ itself increases with $\tau$ and hence, with $U$, $U$ is defined self-consistently. Combining $d_{\rm el}=c\tau$ with $\tau=\tau_0\exp(U/T)$ and noting that $a\approx c\tau_0$, we write

\begin{equation}
d_{\rm el}=a\exp\left(\frac{U}{T}\right)
\label{41}
\end{equation}

\noindent Putting this in Eq. (\ref{4}), we find

\begin{equation}
U=\frac{AT}{T-T_0}
\label{5}
\end{equation}

\noindent where $T_0=3qp_0$ and $A=U_0-T_0\ln\frac{d_0}{2a}$. From Eq. (\ref{5}), the VFT law follows.

In this theory, the origin of the VFT law is the increase of $d_{\rm el}$ on lowering the temperature (see Eqs. (\ref{2}) and (\ref{4})). This increase results in a larger number of LREs that elastically interact with a given event, increasing its activation barrier. The transition from the VFT to the Arrhenius law takes place in the limit of small $d_{\rm el}$ at high temperature. In this case, Eq. (\ref{4}) gives $U=U_0$, i.e. $U$ becomes non-cooperative and temperature-independent. This gives Arrhenius relaxation: $\tau=\tau_0\exp(U_0/T)$.

$T_0=3qp_0$ can be roughly estimated by recalling that $p_0$ is created by local volume increase due to anharmonicity. Because this increase gives rise to macroscopic thermal expansion, $p_0$ can be estimated as $p_0=\alpha BT_m$, where $B$ is bulk modulus, $\alpha$ is the coefficient of thermal expansion and $T_m$ is the temperature of local motion during a LRE, which is on the order of melting temperature. Taking $q\approx 1$ \AA$^3$ and typical liquid values of $B\approx 10$ GPa and $\alpha=10^{-4}-10^{-3}$ K$^{-1}$, we find $T_0=(0.1-1)T_m$, in order-of-magnitude agreement with experimental values \cite{casa}. Hence, $T_0$ in the derived VFT law is related to liquid parameters that are physically sensible.

\section{Effect of volume}

In addition to temperature, our theory also predicts the dependence of $\tau$ on volume or pressure. According to recent experiments, $\tau$ is a function of both temperature and volume, although temperature has a larger overall effect \cite{pr1}. In our theory, pressure has two effects: first, $q$ becomes larger by the amount equal to the decrease of the cage volume due to external pressure $P$, so that the new $q^{\prime}=q+q_c\frac{P}{B}$, where $q_c$ is the initial cage volume. Second, the activation barrier $U$ in Eq. (\ref{4}), and hence the VFT parameter $A$, acquire an additional term $q^{\prime}P$ due to the increased work against the external pressure.

An interesting prediction is the dependence of $T_0$ on pressure: because $T_0=3qp_0$ (see Eq. (\ref{5})), its value under pressure, $T_0^{\prime}$, is $T_0^{\prime}=3p_0q^{\prime}=3p_0\left(q+q_c\frac{P}{B}\right)=T_0+3p_0q_c\frac{P}{B}$. For small $P$, $B$ increases linearly with pressure: $B=B_0+CP$, where $B_0$ is zero-pressure bulk modulus and $C$ is a constant, giving $T_0^{\prime}=T_0+3p_0q_c\frac{P}{B_0+CP}$. Then, for small $\frac{P}{B_0}$, $T_0^{\prime}=T_0+3p_0q_c\frac{P}{B_0}\left(1-C\frac{P}{B_0}\right)$. This behaviour is observed in the experiments: at small $P$, $T_0$ increases linearly with $P$, with a negative quadratic term appearing at higher $P$ \cite{pr3,pr4}.


\section{Dynamic crossovers}

We now discuss how our theory explains the origin of dynamic crossovers, the important problem of glass transition \cite{angell,dyre,phil,langer,kivel,casa,cross1,roland,cross2,sti,sti1}. Experimentally, there are {\it two} dynamic crossovers in a liquid. The {\it first} crossover is at high temperature, and marks the transition from exponential to slow stretched-exponential dynamics and from Arrhenius to VFT relaxation. The physical origin of this crossover has remained one of the central open questions in the area of glass transition \cite{angell,dyre,phil,langer,kivel,casa}. For various liquids, $\tau$ at the crossover is $\tau\approx 1-30$ ps \cite{cross1,roland}.

In our theory, the origin of this crossover is understood as follows. At high temperature when $\tau=\tau_0$, Eq. (\ref{2}) gives $d_{\rm el}=c\tau_0=a\approx 1$ \AA. Therefore, at high temperature elastic waves from LREs do not propagate beyond the nearest-neighbour distance. This means that $d_{\rm el}\approx 1$ \AA\ is shorter than the distance between two neighbouring LREs, or two adjacent molecular cages, $d_m$, which varies from $d_m\approx 10$ \AA\ in small-molecule to $d_m\approx 100$ \AA\ in large-molecule liquids. Because $d_{\rm el}<d_m$, LREs do not elastically interact and, therefore, relax as independent, resulting in exponential and Arrhenius relaxation. On the other hand, when $d_{\rm el}$ increases to $d_m$ on lowering the temperature, LREs are no longer independent, but start interacting via the induced elastic waves. This interaction gives the VFT law as discussed above and, as we recently showed \cite{our1}, stretched-exponential relaxation.

Therefore, the first crossover corresponds to $d_{\rm el}=d_m$. From Eq. (\ref{2}), $\tau$ at the first crossover, $\tau_1$, is

\begin{equation}
\tau_1=\frac{d_m}{c}=\frac{d_m}{a}\tau_0,
\end{equation}
\noindent giving $\tau_1\approx(10-100)\tau_0\approx 1-10$ ps, consistent with the experimental results.

The {\it second} dynamic crossover is at low temperature, and marks another qualitative change in liquid dynamics \cite{cross2,sti,sti1,roland}. The important change is the crossover from the VFT law to a more Arrhenius behaviour. Starting from the early work in Ref. \cite{macedo}, it was realized that at low temperature the VFT law predicts viscosity and $U$ that are larger than those experimentally measured. Moreover, the experimental $U$ at low temperature becomes temperature-independent, contrary to its continuous increase predicted by the VFT law. The crossover from the VFT law to a more Arrhenius form at low temperature has now been established in a large number of glass-forming liquids \cite{roland,sti,sti1}. The origin of this crossover is not understood at present.

In our theory, the origin of this crossover is as follows. When $d_{\rm el}=L$, where $L$ is system size, all LREs in the system elastically interact. This gives temperature-independent $U\propto \ln(L)$ in Eq. (\ref{4}). Hence, $U$ can not increase due to the increase of $d_{\rm el}$ on lowering the temperature, but due to other effects only (e.g., density increase). As a result, $\tau$ crosses over to a more Arrhenius form.

Hence, the second crossover corresponds to $d_{\rm el}=L$. From Eq. (\ref{2}), $\tau$ at the second crossover, $\tau_2$, is

\begin{equation}
\tau_2=\frac{L}{c}=\frac{L}{a}\tau_0
\end{equation}
\noindent Taking a typical experimental value of $L$ in the range of 0.1--10 mm, we find $\tau_2\approx 10^{-7}-10^{-5}$ s. This is in good agreement with $\tau$ of the crossover from the VFT to a more Arrhenius relaxation seen experimentally \cite{sti1}. In addition, $\tau_2$ agrees well with the experimental $\tau$ at which other important liquid properties show a crossover and undergo qualitative changes \cite{cross2}.

Therefore, the two dynamic crossovers originate in our theory in a simple and physically transparent way. Derived solely from the definition of $d_{\rm el}$ in Eq. (\ref{2}), $\tau_1=d_m/c$ and $\tau_2=L/c$ give good agreement with the experiments, without using adjustable parameters. This lends support to the proposed theory of glass transition.

Importantly, the origin of the second dynamic crossover in this theory allows us to reconcile what is perhaps the main ongoing controversy surrounding glass transition, that of divergence \cite{angell,dyre,langer,nat,kivel}. Formally, $\tau$ in the VFT law diverges at $T_0$. Although $T_0$ is always smaller than $T_g$, it is natural to ask what the physical significance of $T_0$ is. Starting from early theories of glass transition, $T_0$ has been associated with a phase transition into a state of zero configurational entropy termed ``ideal glass'' \cite{dyre,langer,nat,kivel}. Confounded by a number of problems \cite{dyre}, this approach was followed by other divergence scenarios of glass transition that were based on the existence of an underlying or avoided phase transition \cite{nat}. However, the nature of the phase transition and the second phase itself remain unclear, primarily because no long-range order or any other appreciable structural changes appear on cooling \cite{dyre,kivel}. Moreover, as recently emphasized, there are no experimentally observed signs of divergence \cite{nat}. These issues continue to fuel the current debate about whether divergence exists and if so, what it means. The problem is often formulated as whether glass transition is a thermodynamic or dynamic phenomenon \cite{dyre,langer,nat,kivel}.

In our theory, this controversy is reconciled as follows. Combining Eqs. (\ref{41}) and Eq. (\ref{5}), we write

\begin{equation}
d_{\rm el}=a\exp\left(\frac{A}{T-T_0}\right)
\end{equation}
\noindent When $T$ approaches $T_0$, $d_{\rm el}$ quickly exceeds any finite size of the system. At this point, $\tau$ crosses over to more Arrhenius as discussed above, moving the divergence to zero temperature. Therefore, $\tau$ does not diverge at $T_0$.

We note that in addition to a seeming divergence of $\tau$ at $T_0$, approaches to glass transition based on thermodynamics and phase transitions have also been stimulated by the experimental changes of heat capacity, compressibility and other properties at $T_g$ \cite{angell,dyre,langer}. As we recently discussed \cite{heat}, these changes can be understood as a natural signature of $T_g$ insofar as $T_g$ is defined by the freezing of LREs at the experimental time scale. In other words, the observed changes are related to a liquid falling out of equilibrium at $T_g$, rather than to thermodynamics or a phase transition \cite{heat}.

\section{Fragility}

Our theory readily explains the origin of liquid fragility \cite{angell,dyre,langer,kivel,angell1}, a widely debated subject in the area of
glass transition. Such an explanation was offered in our recent paper \cite{our2}, which we briefly recall below.

Lets consider two extreme cases of ``strong'' SiO$_2$ and ``fragile'' {\it o}-terphenyl (see Fig. (2)) and calculate $d_{\rm el}$ at the highest measured temperature, $T_h$. From Fig. (2), $\tau(T_h)$ is approximately 10$^{-7}$ s and 10$^{-12}$ s for SiO$_2$ and {\it o}-terphenyl,
respectively. Taking $c\approx 1000$ m/s, $d_{\rm el}(T_h)$ is about 0.1 mm for SiO$_2$ and 1 nm for {\it o}-terphenyl. Hence, $d_{\rm el}$ for {\it o}-terphenyl grows from microscopic to macroscopic values on lowering the temperature. This gives large increase of cooperativity and $U$ in Eq. (\ref{4}),
i.e. fragile behaviour. On the other hand, $d_{\rm el}$ for SiO$_2$ approaches system size at $T_h$ already,
leaving little room for the cooperativity and $U$ to grow on lowering the temperature (see Eq. (\ref{4})). This gives strong behaviour.

\begin{figure}
\begin{center}
\rotatebox{-90}{\scalebox{0.55}{\includegraphics{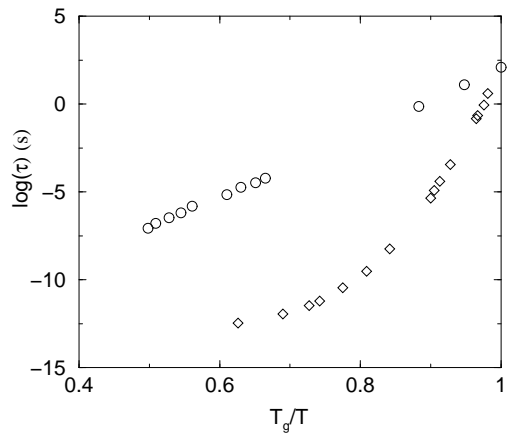}}}
\end{center}
\caption{$\tau$ as a function of $T_g/T$ for SiO$_2$ ($\circ$) and {\it o}-terphenyl ($\diamond$). $\tau$ is calculated from $\eta=G_{\infty}\tau$, where
$G_{\infty}\approx 10$ GPa. The data of $\eta$ are from Ref. \cite{angell1}.}
\end{figure}

In this theory, liquids that have large intrinsic activation barrier and large viscosity ($\tau$) also have large $d_{\rm el}$, even at high 
temperature. This gives little room for $U$ to grow, resulting in strong behaviour. On the other hand, liquids that have small viscosity 
at high temperature and, consequently, small $d_{\rm el}$, show large increase of $U$ on lowering the temperature, i.e. are fragile. 
An interesting prediction of our theory is that if the measurements are extended to higher temperature so that $\tau$ and $d_{\rm el}$ are small,
liquids will become more fragile. Note that there is no data for strong liquids in the range of small $\tau$ \cite{angell1}
(see also, Fig. (2)) due to high melting points, and the evidence for Arrhenius behaviour comes only from the range where $\tau$ is large.

In our theory, fragility can be quantitatively related to other system properties. Fragility is quantified by parameter
$D=\frac{A}{T_0}$ \cite{angell1}; the larger $D$ the smaller fragility. In our theory,
$A=U_0-T_0\ln\frac{d_0}{2a}$ (see Eq. (\ref{5}) and below), giving $D=\frac{U_0}{T_0}-\ln\frac{d_0}{2a}$. Therefore, fragility is predicted to
decrease with the scaled high-temperature activation barrier $U_0$. This is consistent with the experimental data showing the decrease of
fragility with $\frac{U_0}{T_g}$ \cite{novikov}.

\section{Comment on DCS}

We make two remarks about DCS. First, if the frequency of compressive wave fronts arriving at the centre in Fig. 1b is large enough, the atoms in the central cage may, at first glance, appear to be under an effective static stress. The necessary condition for this is that the average time difference between any two compressive arriving wave fronts, $\Delta t$, is shorter than the minimal time set by the elementary vibration period $\tau_0$: $\Delta t<\tau_0$. As discussed above, the central region is affected by all events inside the sphere of radius $d_{\rm el}=c\tau$, each contributing once. Hence, there are $N_{\tau}=\frac{(c\tau)^3}{(d_0/2)^3}$ contributing local relaxing regions inside the sphere of radius $c\tau$. The central event relaxes during time $\tau$, hence the average time between the wave fronts arriving at the centre is ${\Delta t}=\frac{\tau}{N_{\tau}}=\frac{(d_0/2)^3}{c^3\tau^2}$. Because $c\approx a/\tau_0$, the condition ${\Delta t}<\tau_0$ gives $\tau>\tau_0\left(\frac{d_0}{2a}\right)^{\frac{3}{2}}$. Denoting

\begin{equation}
\tau_{min}=\tau_0\left(\frac{d_0}{2a}\right)^{\frac{3}{2}}
\end{equation}
\noindent we find that $\tau_{min}\approx$1 ps, i.e. is very short.

As discussed in the previous section, $d_{\rm el}$ exceeds the experimental system size $L$ above $T_g$. In this case, we should substitute $c\tau$ for $L$, and the condition ${\Delta t}<\tau_0$ gives $\tau<\tau_0\left(\frac{L}{d_0}\right)^3$. Denoting

\begin{equation}
\tau_{max}=\tau_0\left(\frac{L}{d_0}\right)^3
\end{equation}
\noindent and using a typical value of $L$ of 1 mm, we find $\tau_{max}\approx 10^5$ s. Because $\tau_{max}\gg\tau(T_g)$, the condition ${\Delta t}<\tau_0$ is also satisfied when $d_{\rm el}>L$ for a macroscopic system.

Therefore, the necessary condition for the cage atoms to be under the static stress (${\Delta t}<\tau_0$) is fulfilled in a very wide range of $\tau$, including glass transformation range. However, the actual value of this stress, $p_s$, vanishes for $\tau\gg\tau_0$. Indeed, the contributions from arriving compressive wave fronts to the static stress can be summed on the time scale of $\tau_0$ only, for two reasons. First, wave fronts from remote LREs add up only if they arrive simultaneously, or in practice during the elementary time period $\tau_0$. Second, the duration of the passing wave front is also $\tau_0$, because the lifetime of the remote LRE that creates the wave is about $\tau_0$. Therefore, $p_s$ can be calculated as

\begin{equation}
p_s=\rho_{\tau_0}\int\limits_{d_0/2}^{d_{\rm el}} 4\pi r^2 p_i (r) {\rm d}r
\label{ps}
\end{equation}

\noindent where $\rho_{\tau_0}$ is the density of those remote local regions that give rise to the waves passing through the centre during time $\tau_0$.

The number of the wave fronts that pass through the centre during time $\tau_0$ is $N_{\tau_0}=\frac{\tau_0}{\Delta t}$. Combining this with ${\Delta t}=\frac{\tau}{N_{\tau}}$ from above gives $N_{\tau_0}=\frac{\tau_0}{\tau}{N_{\tau}}$. Hence, $\rho_{\tau_0}=\frac{\tau_0}{\tau}\rho$, where $\rho=\frac{6}{\pi d_0^3}$ is the density of local relaxing regions introduced in Eq. (\ref{3}). Using $\rho_{\tau_0}$ in Eq. (\ref{ps}) and recalling that $d_{\rm el}=c\tau=\frac{a\tau}{\tau_0}$ and $p_i(r)=p_0\left(\frac{d_0}{2r}\right)^3$, we write

\begin{equation}
p_s=3p_0\frac{\tau_0}{\tau}\left(\ln\frac{\tau}{\tau_0}+\ln\frac{2a}{d_0}\right)
\end{equation}

Therefore, $p_s$ vanishes for $\tau\gg\tau_0$, i.e. for $\tau$ at which $d_{\rm el}$ and elastic interactions between LREs become appreciable in the first place (see the previous section). As a result, DCS does {\it not} contribute to liquid internal pressure or equilibrium volume. Physically, this is because the volume of local regions that contribute to $p_s$ is sparse.

We note that the vanishing of $p_s$ is in contrast to the behaviour of $U$ which increases with $\tau$ and $d_{\rm el}$ (see Eq. (\ref{4})). $U$ is given by the total work to be performed against cage expansion \cite{frenkel}. Because each LRE compresses the central cage, this work, being an extensive quantity, is the sum over all pulses that arrive during time $\tau$ from within the sphere of radius $d_{\rm el}$. Therefore, $U$ increases with the number of contributing LREs and $d_{\rm el}$.

The second remark about DCS concerns Eq. (\ref{1}). After the initial expansion due to a LRE, a remote local region relaxes back to its original equilibrium volume, sending a dilatational wave to the centre. However, the dynamic dilatational stress, $p_d$, and the
dynamic compressive stress, $p$, do not oscillate around an equilibrium value as in a harmonic wave, but are set by processes that operate on different time scales (or, equivalently, length scales, see below), making propagating density variations essentially anharmonic. Recall that $p$ is due to fast expansion originating from anharmonic forces that initially appear between several (3--5) ``hot'' atoms of the remote LRE during the
elementary vibration period $\tau_0\approx 0.1$ ps. As the expanding wave propagates, fast dynamic compression is created between the two adjacent
atomic layers when one layer is already displaced during $\tau_0$ but the other is not.

On the other hand, dilatation that sets $p_d$ and originates from cooling and contraction of the remote relaxing region is a slower process for the
following reason. Let $t_d$ be the time during which $p_d$ exists. $t_d$ is set by the time during which increased interatomic separations
in the expanded remote region, $\Delta r$, return to their original (pre-LRE) values, $\Delta r_0$. This process involves the number of atoms
that is necessarily larger than the number of initially hot atoms, because these hot atoms interact with their neighbours. As a result of this
interaction, large $\Delta r$ between the initially hot atoms are distributed over a larger region. This is accompanied by a gradual reduction of $\Delta r$. As thermalization of the local region proceeds and $\Delta r=\Delta r_0$ is established for all atoms involved, the
dilatation of the relaxing region is complete. Therefore, $t_d$ is set by the time of thermalization of atoms affected by the LRE motion. This thermalization involves at least the nearest neighbours of the initially hot atoms, hence $t_d$ is the time of thermalization of the region whose
size is at least equal to the cage size $d_0$. Therefore, the lower limit of $t_d$ can be estimated as $\frac{d_0}{c}\approx$ 1 ps, because
thermalization can not proceed faster than the phonon speed, and we find $t_d\ge$1 ps. Hence, $t_d\gg\tau_0$.

Interestingly, a similar effect is observed in molecular dynamics simulations of radiation damage.
Here, a hot radiation cascade created by an energetic ion elastically deforms the surrounding lattice. It is found that fast initial
expansion of the lattice is followed by considerably slower contraction due to finite thermal conductivity \cite{radi}.

$t_d\gg\tau_0$ means that $p_d\ll p$. A mechanical analogy of this effect is compression of a spring due to a fast compressive force applied to one end, followed by slow motion in the opposite direction during which the spring length hardly changes.


$\frac{p_d}{p}$ can be related to $\frac{\tau_0}{t_d}$ as follows. Lets consider that the boundary atom in the central cage is pushed towards the central atom by force $f$ due to compressive stress $p$, followed by the motion in the opposite direction caused by force $f_d$ due to dilatational stress $p_d$ that arrives later. Let $l$ be forward and reverse displacement of the boundary atom due to $p$ and $p_d$, respectively.
Assuming constant acceleration, $f=\frac{2ml}{\tau_0^2}$ and $f_d=\frac{2ml}{t_d^2}$, where $m$ is the atomic mass. Then, $\frac{p_d}{p}=\frac{f_d}{f}=\left(\frac{\tau_0}{t_d}\right)^2$. Hence, $p_d\ll p$ because $\tau_0\ll t_d$.

$\frac{p_d}{p}$ can also be written as the ratio of the short-range and medium-range order distances, $\frac{a}{d_0}$. Using $t_d\ge \frac{d_0}{c}$ and $c=\frac{a}{\tau_0}$, $\frac{p_d}{p}=\left(\frac{\tau_0}{t_d}\right)^2<\left(\frac{a}{d_0}\right)^2$. Hence, $p_d\ll p$ because $a\ll d_0$. Physically, this result has the following meaning. A statement equivalent to $\tau_0\ll t_d$ is the assertion that the wavelength of the initial compressive wave is comparable with interatomic separations in the central cage (because this wave is created by fast expansion of the remote cage), whereas the wavelength of the following dilatational wave is larger than the cage size due to slow contraction. Hence, interatomic separations in the central cage decrease in the first case, whereas the cage moves as a whole in the second, resulting in $p_d\ll p$.

From $p_d\ll p$, $qp_d\ll qp$ follows, i.e. the negative work due to $p_d$ is small, and can be ignored in Eq. (\ref{1}).

\section{Summary}

In summary, we proposed that the properties of glass transition can be understood on the basis of elastic waves in a liquid. Elastic waves originating from atomic jumps in a liquid create a dynamic compressive stress, which slows down relaxation. The increase of $d_{\rm el}$ on lowering the temperature gives the VFT law. In addition to temperature, we also predicted the effect of volume on $\tau$. In the proposed theory, we discussed the origin of dynamic crossovers, the absence of divergence of $\tau$ at $T_0$, the effect of volume on $\tau$ and the origin of liquid fragility.

We are grateful to R. Casalini, M. T. Dove, V. Heine, A. Navrotsky and C. M. Roland for discussions, and to EPSRC and RFBR for support.

\appendix*
\section{Appendix}

In this Appendix, we establish the equivalence of $d_{\rm el}$ introduced in Eq. (\ref{2}) and the result from the viscoelastic theory of Frenkel \cite{frenkel}. The relevant part of Frenkel's discussion starts with the modification of elasticity equations due to relaxation process in a liquid. Consider, for example, the relationship between shear stress $P$ and shear strain $s$: $P=2Gs$, where $G$ is shear modulus. In the presence of relaxation process the strain includes an extra displacement due to viscous response, and the total strain is written according to Maxwell interpolation as

\begin{equation}
\frac{ds}{dt}=\frac{P}{2\eta}+\frac{1}{2G}\frac{dP}{dt}
\label{a1}
\end{equation}
\noindent

Introducing the operator

\begin{equation}
A=1+\tau\frac{d}{dt},
\label{a2}
\end{equation}
\noindent where $\tau=\eta/G$, Eq. (\ref{a1}) can be written as $\frac{ds}{dt}=\frac{1}{2\eta}AP$. If $A^{-1}$ is the reciprocal operator to $A$, $P=2\eta A^{-1}\frac{ds}{dt}$. Because $\frac{d}{dt}=\frac{A-1}{\tau}$ from Eq. (\ref{a2}), $P=2G(1-A^{-1})s$. Comparing this with $P=2Gs$, we find that the presence of relaxation process is equivalent to the substitution of $G$ by the operator $M=G(1-A^{-1})$.

Consider the propagation of the wave of $P$ and $s$ with time dependence $\exp(i\omega t)$. Differentiation gives multiplication by $i\omega$. Then, $A=1+i\omega\tau$, and $M$ is:

\begin{equation}
M=\frac{G}{1+\frac{1}{i\omega\tau}}
\label{a4}
\end{equation}

If $M=R\exp(i\phi)$, the inverse complex velocity is $\frac{1}{v}=\sqrt\frac{\rho}{M}=\sqrt\frac{\rho}{R}(\cos\frac{\phi}{2}-i\sin\frac{\phi}{2})$, where $\rho$ is density. $P$ and $s$ depend on time and position $x$ as $f=\exp(i\omega(t-x/v))$. Using the above expression for $v$, $f=\exp(i\omega t)\exp(-ikx)\exp(-\beta x)$, where $k=\omega\sqrt\frac{\rho}{R}\cos\frac{\phi}{2}$ and absorbtion coefficient $\beta=\omega\sqrt\frac{\rho}{R}\sin\frac{\phi}{2}$. Combining the last two expressions, we write $\beta=\frac{2\pi\tan\frac{\phi}{2}}{\lambda}$, where $\lambda=\frac{2\pi}{k}$ is the wavelength.

From Eq. (\ref{a4}), $\tan\phi=\frac{1}{\omega\tau}$. For high-frequency waves $\omega\tau\gg 1$, $\tan{\phi}\approx\phi=\frac{1}{\omega\tau}$, giving $\beta=\frac{\pi}{\lambda\omega\tau}$. This is Frenkel's result \cite{frenkel}. Essentially the same expression is obtained for high-frequency longitudinal waves \cite{frenkel}. Here, Eqs. (\ref{a1}-\ref{a4}) are modified to include a finite zero-frequency bulk modulus.

Lets introduce the dissipation length $d=1/\beta$ so that $f\propto\exp(-x/d)$. Then, $d=\frac{\lambda\omega\tau}{\pi}$. Because $\omega=\frac{2\pi c}{\lambda}$, $d=2c\tau$ (if $\omega$ is close to Debye frequency, the last two formulas are correct approximately). Therefore, Frenkel's theory gives $d\approx c\tau$ as in Eq. (\ref{2}).

We note that essential to Eq. (\ref{2}) is the solid-like character of LRE-induced waves: because $\omega\gg 1/\tau$ holds true for these waves, they propagate in elastic equilibrium. Our derivation of Eq. (\ref{2}) was based on this assertion, because we considered how elastic waves in a solid are modified in the presence of LREs. Hence, the condition $\omega\gg 1/\tau$ is used in both our and Frenkel's derivation above. Importantly, $d_{\rm el}=c\tau$ {\it increases} with $\tau$ in this regime of wave propagation.

A different situation arises when $\omega\ll 1/\tau$. More frequently discussed in the literature, this case corresponds not to elastic, but to hydrodynamic equilibrium. When $\omega\tau\ll 1$, Frenkel's theory gives for shear waves $\phi=\frac{\pi}{2}$ and $d=\frac{\lambda}{2\pi}$. Different from the high-frequency case, this means that low-frequency shear waves are not propagating (because they are dissipated over the distance comparable to the wavelength), a result that is also known from hydrodynamics \cite{hydro}. For low-frequency $\omega\tau\ll 1$ longitudinal waves, Frenkel's theory gives $d$ on the order of $\frac{\lambda}{{\omega}{\tau}}$ \cite{frenkel}. Here, $d$ {\it decreases} with $\tau$, in agreement with hydrodynamics \cite{hydro}, but in contrast to the high-frequency case due to a different regime of wave propagation.

\end{document}